%% file: main.tex
\documentclass[fp, seceq]{arxiv}
\usepackage{bm}
\usepackage[dvipdfmx]{graphicx}
\usepackage{algorithm}
\usepackage{algpseudocode}

\title{
  Community Detection Algorithm Combining Stochastic Block Model and Attribute Data Clustering
}

\author{
  Shun Kataoka$^1$\thanks{xkataoka@smapip.is.tohoku.ac.jp}, 
  Takuto Kobayashi$^1$, 
  Muneki Yasuda$^2$,
  and Kazuyuki Tanaka$^1$
}

\inst{
  $^1$Graduate School of Information Science, Tohoku University.\\
  $^2$Graduate School of Science and Engineering, Yamagata University.
}

\abst{
  We propose a new algorithm to detect the community structure in a network that utilizes both the network structure and vertex attribute data.
  Suppose we have the network structure together with the vertex attribute data, that is, the information assigned to each vertex associated with the community to which it belongs.
  The problem addressed this paper is the detection of the community structure from the information of both the network structure and the vertex attribute data.
  Our approach is based on the Bayesian approach that models the posterior probability distribution of the community labels.
  The detection of the community structure in our method is achieved by using belief propagation and an EM algorithm.
  We numerically verified the performance of our method using computer-generated networks and real-world networks.
}

\begin{document}
\maketitle

\input{introduction}
\input{relatedWorks}
\input{modelDefinition}
\input{BP}
\input{EM}
\input{exp}
\input{conc}

\section*{Acknowledgement}
This work was partially supported by CREST, Japan Science and Technology Agency, and by JSPS KAKENHI (Grant Numbers 15K20870, 15K00330, and 25280089).

\bibliographystyle{my-jpsj}
\bibliography{stringAbb,ref}

\end{document}

%% file: introduction.tex
\section{Introduction}
The division of a large amount of data into groups is a fundamental and important task for data analysis and understanding in various scientific fields, such as image processing\cite{GonzalezWoodsBook2007}, machine learning\cite{MurphyBook2012}, and bioinformatics\cite{PolanskiKimmelBook2007}.
A simplified representation of the data grouped according to a certain similarity provides a considerable amount of meaningful information that is easier to understand and analyze than the original information; e.g., segments of an image constitute a useful description for detecting a certain object in the image.

In the field of complex networks\cite{NewmanBook2010}, the task of dividing a network into subnetworks for network simplification, analysis, and understanding is called community detection\cite{FortunatoPhysRep2010}.
Unfolding the community structures in networks is an important problem for network analysis, because vertices in the same community tend to have the same functional properties and it facilitates the discovery of the hierarchical structures that characterize many existing networks, e.g. the company, department, and division relationship in a human network of business people.
In the context of complex networks, this task seems to be recognized as a structure-based partition problem, the objective of which is to divide the network into groups of vertices such that the connections between the vertices in the same group are denser than the connections to the vertices of the other groups.
However, vertices in many existing networks have attribute data associated with the community to which it belong, such as age, wealth, or affiliation in a human society network.
Therefore, a community detection method that considers both vertex attribute data and network structure can be expected to provide more salient results than previous methods that considered only network structure.

In this paper, we propose a new community detection method based on probabilistic modeling of networks with attribute data.
We use a Bayesian approach to express a posterior probability distribution of the community labels.
Our method can be regarded as a natural extension of the community detection method that uses a message passing method\cite{HastingsPRE2006, DecelleKrzakalaMooreZdeborovaPRL2011, DecelleKrzakalaMooreZdeborovaPRE2011} combined with a data clustering method using a mixture of Gaussian distribution\cite{MurphyBook2012} or an extension of the method proposed by Zanghi et al.\cite{ZanghiVolantAmbroisePattRecogLett2010} from the perspective of the cluster variation method\cite{PelizzolaJPhysA2005, YedidiaFreemanWeissIEEE2005}.
%Due to the probability expression of the network model these extension can be easily achieved.

The remainder of this paper is organized as follows.
In section \ref{sec::relatedWorks}, we describe the positioning of our work as compared with the related work.
In section \ref{sec::modelDef}, we define a probability model of community labels that considers both network structure and vertex attribute data.
In section \ref{sec::BP}, we derive a community inference method based on the posterior probability distribution described in section \ref{sec::modelDef}.
In section \ref{sec::EM}, we describe a framework for determining the model parameters in the posterior probability distribution.
In section \ref{sec::Exp}, we give numerical results that verify the performance of our proposed method when applied to computer-generated and real-world networks.
Finally in section \ref{sec::conclude}, we present our concluding remarks.

%% file: relatedWorks.tex
\section{Related Work} \label{sec::relatedWorks}
The research of detecting communities in networks has a long history and had been developed in the field of sociology\cite{BurtSocForces1976} before the dawn of the complex network\cite{WattsStrogatzNature1998,BarabasiAlbertScience1999}.
However, community detection is still an active research topic in network science because of its wide application in fields including social science, computer science, biology, etc., and many methods to detect communities were proposed at the beginning of the 21st century.
In the early period of complex network research, Girvan and Newman proposed a detection method that uses betweenness centrality\cite{GirvanNewmanNAS2002}.
Then, Newman also proposed a well known modularity optimization method that optimizes the modularity using a greedy optimization method to detect the community structure in networks\cite{NewmanPRE2004,ClausetNewmanMoorePRE2004}.
As a variant of the modularity optimization method, Blondel et al. proposed the Louvain algorithm, which iteratively optimizes the modularity by creating a new network the vertices of which are communities\cite{BlondelGuillaumeLambitteLefebureJStatMech2008}.
From the statistical mechanical perspective, Reichardt and Bornholdt proposed a community detection method that is aimed to find the ground state of the Hamiltonian by using simulated annealing\cite{ReichardtBornholdtPRL2004, ReichardtBornholdtPRE2006}.
As a probabilistic approach, Hasting proposed a message passing method based on the stochastic block model\cite{HastingsPRE2006}, and then, his method was extended and analyzed in depth by Decelle et al.\cite{DecelleKrzakalaMooreZdeborovaPRL2011, DecelleKrzakalaMooreZdeborovaPRE2011}.

The common point of most community detection algorithms, including the methods mentioned above, is that only the network structures are taken into account to find the communities.
However, as mentioned in the previous section, networks in the real world contain other information related to the community structure and such information can be freely utilized to detect the communities if it can be determined that a relationship exits between the community structure and such information.

Recently, community detection methods that consider vertex assigned information, called attribute data, and the network structure have been proposed by several researchers.
Most of these methods were formulated by extending the previous community detection methods that consider only network structure or traditional data clustering methods, such as k-means clustering and k-medoids clustering\cite{KaufmanRousseeuwBook1990}.
Dang and Viennet proposed two modularity optimization-based methods taht are extensions of the Louvain algorithm, where the effect of the vertex attribute data was considered by defining a new cost function or creating a new network structure\cite{DangViennetICDS2012}.
Zhou et al. proposed a detection method that extends the k-medoids clustering method by defining the distance for measuring the closeness on an attribute in a considered network\cite{ZhouChengYuVLDB2009}.
As a different kind of approach based on the probabilistic model that considers both the network structure and vertex attribute data, Yang et al. proposed a probabilistic method that allows the detection of overlapping communities, where communities are detected by thresholding the community membership parameters after maximizing its likelihood function\cite{YangMcAuleyLeskovecICDM2013}.

The importance of these methods is in that they allow the discrete and textual vertex attribute data to be utilized for detecting the community structure in a network by virtue of the extensions described above.
On the other hand, we can consider an additional type of attribute data, i.e., continuous vertex attribute data, which can be regarded as feature quantities for the community detection problem, similar to the feature quantities for the pattern recognition problem\cite{TheodoridisKoutroumbasBook2009}, and which are extracted from the raw data such as the textual data utilized by the above methods in the expectation of increasing the detection accuracy.
In this direction, Zanghi et al. proposed a community detection method that considers both the network structure and vertex attribute data based on a stochastic block model and mixture of Gaussian distribution\cite{ZanghiVolantAmbroisePattRecogLett2010}.
Furthermore, naive mean field approximation\cite{ParisiBook1988} is used to infer the community labels in their method.

In this paper, we derive a message passing algorithm to find the community structure that considers both the network structure and vertex attribute data by using belief propagation\cite{PearlBook1988}.
Our work is a direct extension of the research of Zanghi et al. from the perspective of the cluster variation method and can be regarded as an extension of the previous message passing method that considers only network structure by including a Gaussian mixture data clustering method.
As in previous studies\cite{DecelleKrzakalaMooreZdeborovaPRL2011, DecelleKrzakalaMooreZdeborovaPRE2011, ZanghiVolantAmbroisePattRecogLett2010}, we utilize the stochastic block model to express the posterior probability distribution of the community labels and an EM algorithm\cite{DempsterLairdRubinJRSS1977} to determine the parameters of the posterior probability distribution.
Although there is another similar method that uses the variational Bayesian approach\cite{JordanGhahramaniJaakkolaML1999} on the stochastic block model that considers vertex attribute data proposed by Xu et al.\cite{XuKeWangCheng2ACMT2014}, the authors adopted raw discrete data directly as the attribute data and did not target the extracted feature quantity as the original direction.
However, the combination of our approach and the variational Bayesian approach\cite{JordanGhahramaniJaakkolaML1999} constitutes a very interesting extension and we leave this task for future work.

%% file: modelDefinition.tex
\section{Model Definition} \label{sec::modelDef}
In this section, we define the posterior probability distribution for community detection using vertex attribute data.
Let $V$ and $E$ be the set of vertices and set of edges of the observed undirected network, respectively.
We define $A$ as an adjacency matrix of the observed network, the $ij$ elemrnt of which is expressed as
\begin{align}
  A_{ij}
  =
  \begin{cases}
    1, & ij \in E \\
    0, & \mbox{otherwise}
  \end{cases}
\end{align}
and assign attribute value $d_i$, which is a real value, to each vertex $i \in V$.
Suppose that the network has $L_{\mbox{\scriptsize max}}$ communities and $x_i \in L = \left\{ 1, \dots, L_{\mbox{\scriptsize max}} \right\}$ is a random variable denoting the community label of vertex $i \in V$.

In the Bayesian point of view, the detection of the community labels is inferred by using the posterior probability distribution $P \left( \bm{x} \middle| A, \bm{d} \right)$ expressed as 
\begin{align}
  P \left( \bm{x} \ \middle| \  A, \bm{d} \right)
  = & 
  \frac{
    P \left( A, \bm{d} \ \middle| \  \bm{x} \right) P \left( \bm{x} \right)
  }{
    \sum_{\bm{x}} P \left( A, \bm{d} \ \middle| \ \bm{x} \right) P \left( \bm{x} \right)
  } \nonumber \\
  = & 
  \frac{
    P \left( A \ \middle| \ \bm{x} \right) P \left( \bm{d} \ \middle| \ \bm{x} \right) P \left( \bm{x} \right)
  }{
    \sum_{\bm{x}} P \left( A \ \middle| \ \bm{x} \right) P \left( \bm{d} \ \middle| \ \bm{x} \right) P \left( \bm{x} \right)
  }, \label{eq::bayes}
\end{align}
where $\bm{x} = \left\{ x_i \ \middle| \ i \in V \right\}$ and $\bm{d} = \left\{ d_i \ \middle| \ i \in V \right\}$ are sets of random variables and attribute data, respectively.
The summation $\sum_{ \bm{x} }$ denotes the multiple summations over all the possible configurations of $\bm{x}$.
We assume that the adjacency matrix $A$ is conditionally independent of attribute data $\bm{d}$ given $\bm{x}$ in the last line of Eq. (\ref{eq::bayes}) for simplicity.

To define a concrete form of $P \left( \bm{x} \ \middle| \ A, \bm{d} \right)$, we assume that the prior probability distribution $P \left( \bm{x} \right)$ is expressed as 
\begin{align}
  P \left( \bm{x}; \bm{\gamma} \right)
  = 
  \prod_{ i \in V } \prod_{ l \in L } \gamma_l^{ \delta \left( x_i, l \right) }, \label{eq::prior}
\end{align}
where $\bm{\gamma} = \left\{ \gamma_l \in [0, 1] \ \middle| \ l \in L \right\}$ is a set of parameters and normalized as 
\begin{align}
  \sum_{ l \in L } \gamma_l = 1 \label{eq::gamma_normalize}
\end{align}
and $\delta \left( a, b \right)$ is the Kronecker delta.
In this paper, we express the parameters of the function by its arguments after the semicolon as in Eq. (\ref{eq::prior}).
The parameter $\gamma_l$ is a prior probability with which the label $l$ is assigned to each vertex.
We assume that the observed network structure $E$ is generated by connecting vertex $i$ and vertex $j$ according to their labels $x_i$ and $x_j$ with probability $\Gamma_{ls} \in [0, 1]$:
\begin{align}
  P \left( A \ \middle| \ \bm{x}; \Gamma \right)
  = 
  \prod_{ ij \in I } \prod_{ l \in L } \prod_{ s \in L } 
  \left[ \Gamma_{ls}^{A_{ij}} \left( 1 - \Gamma_{ls} \right)^{1 - A_{ij}} \right]^{\delta \left( x_i, l \right) \delta \left( x_j, s \right)}, \label{eq::condi1}
\end{align}
where $\Gamma = \left\{ \Gamma_{ls} \ \middle| \ l, s \in L, \Gamma_{ls} = \Gamma_{sl} \right\}$ is a set of parameters $\Gamma_{ls}$.
$I = \left\{ ij \ \middle| \ i, j \in V, i < j \right\}$ is a set of all the distinct pairs of the vertices.
This conditional distribution asserts that vertex $i$ with label $l$ and vertex $j$ with label $s$ are connected with probability $\Gamma_{ls}$.
The conditional probability density function $P \left( \bm{d} \ \middle| \ \bm{x} \right)$ is defined as
\begin{align}
  P \left( \bm{d} \ \middle| \ \bm{x} ; \Theta \right)
  = 
  \prod_{ i \in V } \prod_{ l \in L } \mathcal{N} \left( d_i; \mu_l, \sigma_l \right)^{\delta \left( x_i, l \right)}, \label{eq::condi2}
\end{align}
where $\Theta = \left\{ \mu_l, \sigma_l \ \middle| \ l \in L \right\}$ and $\mathcal{N} \left( c; \mu, \sigma \right)$ is a normal distribution of mean $\mu$ and variance $\sigma^2$.
This conditional probability density function represents our intuition that vertices belonging to the same community have similar attribute data.
Although we assume that attribute data have a one-dimensional value, the extension to multidimensional attribute data is straightforward.

By substituting Eqs. (\ref{eq::prior}), (\ref{eq::condi1}), and (\ref{eq::condi2}) in Eq. (\ref{eq::bayes}), the posterior distribution $P \left( \bm{x} \ \middle| \ A, \bm{d} \right)$ is expressed as 
\begin{align}
  P \left( \bm{x} \ \middle| \ A, \bm{d}; \bm{\gamma}, \Gamma, \Theta \right)
  \propto 
  \prod_{ i \in V } \phi_i \left( x_i \ \middle| \ d_i; \bm{\gamma}, \Theta \right) 
  \prod_{ ij \in I } \phi_{ij} \left( x_i, x_j \ \middle| \ A; \Gamma \right), \label{eq::posterior}
\end{align}
where 
\begin{align}
  \phi_i \left( x_i \ \middle| \ d_i; \bm{\gamma}, \Theta \right) 
  = 
  \prod_{ l \in L } \left[ \gamma_l \mathcal{N} \left( d_i; \mu_l, \sigma_l \right) \right]^{\delta \left( x_i, l \right)}
\end{align}
and 
\begin{align}
  \phi_{ij} \left( x_i, x_j \ \middle| \ A; \Gamma \right) 
  = 
  \prod_{ l \in L } \prod_{ s \in L } \left[ \Gamma_{ls}^{A_{ij}} \left( 1 - \Gamma_{ls} \right)^{1 - A_{ij}} \right]^{\delta \left( x_i, l \right) \delta \left( x_j, s \right)},
\end{align}
respectively. 
This is the same probability model that Zanghi et al. used in their study\cite{ZanghiVolantAmbroisePattRecogLett2010}.
It is worth noting that the joint probability distribution $\bm{x}$ and $A$
\begin{align}
  P \left( \bm{x}, A ; \bm{\gamma}, \Gamma \right)
  = 
  \prod_{ i \in V } \prod_{ l \in L } \gamma_l^{\delta \left( x_i, l \right)}
  %\times 
  \prod_{ ij \in I } \prod_{ l \in L } \prod_{ s \in L } 
  \left[ \Gamma_{ls}^{A_{ij}} \left( 1 - \Gamma_{ls} \right)^{1 - A_{ij}} \right]^{\delta \left( x_i, l \right) \delta \left( x_j, s \right)} \label{eq::SBM}
\end{align}
obtained from Eqs. (\ref{eq::prior}) and (\ref{eq::condi1}) is the stochastic block model used in previous studies\cite{HastingsPRE2006, DecelleKrzakalaMooreZdeborovaPRL2011, DecelleKrzakalaMooreZdeborovaPRE2011}.
Therefore, the posterior probability distribution $P \left( \bm{x} \ \middle| \ A, \bm{d}; \bm{\gamma}, \Gamma, \Theta \right)$ in Eq. (\ref{eq::posterior}) can be regarded as a straightforward extension of the stochastic block model that considers the vertex attribute data in the Bayesian framework.

%% file: BP.tex
\section{Inference Algorithm based on Belief Propagation} \label{sec::BP}
In this section, we propose an inference algorithm for detecting communities in an observed network from network structure $E$ and attribute data $\bm{d}$.
In our method, community labels are estimated by finding the convergence point of message passing rules of belief propagation\cite{PearlBook1988}.

A standard technique for estimating the community labels from a posterior probability distribution is the MAP estimation method that finds the labels that maximize the posterior probability distribution.
However, MAP estimation is  difficult because finding the estimate labels that maximize $P \left( \bm{x}, A; \bm{\gamma}, \Gamma \right)$ in Eq. (\ref{eq::posterior}) is an NP-hard problem.
Therefore, we adopt belief propagation, which is an approximate inference method that computes the approximate marginal probability distributions of $P \left( \bm{x}, A; \bm{\gamma}, \Gamma \right)$ for vertex $i$ and pair $ij \in I$ denoted by $b_i \left( x_i \right)$ and $b_{ij} \left( x_i, x_j \right)$, respectively, instead of using MAP estimation.
The estimation of community labels is achieved by finding $\hat{\bm{x}} = \left\{ \hat{x}_i \ \middle| \ i \in V \right\}$ such that
\begin{align}
  \hat{x}_i 
  = 
  \arg \max_{x_i} b_i \left( x_i \right) \label{eq::MPM}
\end{align}
for $i \in V$.
This type of estimation method that finds the arguments that maximize the marginal probability distributions is called maximization of the posterior marginals(MPM) estimation.
In the framework of belief propagation, the approximate marginal distribution $b_i \left( x_i \right)$ and $b_{ij} \left( x_i, x_j \right)$ is given by 
\begin{align}
  b_i \left( x_i \right) 
  \propto 
  \phi_i \left( x_i \ \middle| \ d_i; \bm{\gamma}, \Theta \right) \prod_{ k \in V \backslash \left\{ i \right\} } M_{k \rightarrow i} \left( x_i \right) \label{eq::vertexB}
\end{align}
and
\begin{align}
  b_{ij} \left( x_i, x_j \right)
  \propto &
  \phi_{ij} \left( x_i, x_j \ \middle| \ A; \Gamma \right) \nonumber \\
  &\times \left[ \phi_i \left( x_i \ \middle| \ d_i; \bm{\gamma}, \Theta \right) \prod_{k \in V \backslash \left\{ i,j \right\}} M_{k \rightarrow i} \left( x_i \right) \right] \nonumber \\
  &\times \left[ \phi_j \left( x_j \ \middle| \ d_j; \bm{\gamma}, \Theta \right) \prod_{k \in V \backslash \left\{ i,j \right\}} M_{k \rightarrow j} \left( x_j \right) \right], \label{eq::edgeB}
\end{align}
respectively.
$M_{j \rightarrow i} \left( x_i \right)$ in Eqs. (\ref{eq::vertexB}) and (\ref{eq::edgeB}) is a message from vertex $j$ to vertex $i$ and is obtained by the convergence points of the message passing rule
\begin{align}
  M_{j \rightarrow i} \left( x_i \right)
  = 
  \frac{1}{Z_{j \rightarrow i}} \sum_{ x_j } \phi_{ij} \left( x_i, x_j \ \middle|\  A; \Gamma \right) 
  \phi_j \left( x_j \ \middle| \ d_j; \bm{\gamma}, \Theta \right) 
  \left[ \prod_{k \in V \backslash \left\{ i,j \right\}} M_{k \rightarrow j} \left( x_j \right) \right], \label{eq::BPrule1}
\end{align}
where $Z_{j \rightarrow i}$ is a normalization constant.

Community labels are estimated by solving the simultaneous equations in Eq. (\ref{eq::BPrule1}) by means of an iteration method.
However, we need to treat $\left| V \right| \left( \left| V \right| - 1 \right)$ messages to compute marginal distributions $b_i \left( x_i \right)$ in this framework, where $\left| S \right|$ denotes the cardinality of the set $S$.
Therefore, we approximate our message passing rule according to the derivation of the message passing rule in previous studies\cite{HastingsPRE2006, DecelleKrzakalaMooreZdeborovaPRL2011, DecelleKrzakalaMooreZdeborovaPRE2011} to reduce the number of messages to $2 \left| E \right|$ by assuming that the observed network is a large sparse graph, that is, $\left| E \right| = O \left( \left| V \right| \right)$ and $\left| V \right| \gg 1$.
For this approximation, we define the new message from vertex $j$ to vertex $i$ as 
\begin{align}
  m_{j \rightarrow i} \left( x_j \right)
  \propto 
  \phi_j \left( x_j \ \middle| \ d_j; \bm{\gamma}, \Theta \right) 
  \left[ \prod_{k \in V \backslash \left\{ i,j \right\}} M_{k \rightarrow j} \left( x_j \right) \right], \label{eq::newMessage}
\end{align}
By substituting Eq. (\ref{eq::newMessage}) in Eq. (\ref{eq::BPrule1}) and using the relation
\begin{align}
  M_{j \rightarrow i} \left( x_i \right)
  \propto 
  \sum_{ x_j } \phi_{ij} \left( x_i, x_j \ \middle| \ A; \Gamma \right) m_{j \rightarrow i} \left( x_j \right),
\end{align}
we can derive the message passing rules of new messages as
\begin{align}
  m_{j \rightarrow i} \left( x_j \right)
  = 
  \frac{1}{z_{j \rightarrow i}} \phi_j \left( x_j \ \middle| \ d_j; \bm{\gamma}, \Theta \right)
  %\times 
  \prod_{k \in V \backslash \left\{ i,j \right\}} 
    \left[ 
    \sum_{ x_k } \phi_{jk} \left( x_j, x_k \ \middle| \ A; \Gamma \right) m_{k \rightarrow j} \left( x_k \right)
    \right] \label{eq::newMessageTmp},
\end{align}
where $z_{ji}$ is a normalization constant.
The marginal distribution $b_i \left( x_i \right)$ $b_{ij} \left( x_i, x_j \right)$ can be expressed as
\begin{align}
  b_i \left( x_i \right)
 \propto 
 \phi_i \left( x_i \ \middle| \ d_i; \bm{\gamma}, \Theta \right)
 \prod_{ k \in V \backslash \left\{ i \right\} } 
 \left[
   \sum_{ x_k } \phi_{ik} \left( x_i, x_k \ \middle| \ A; \Gamma \right) m_{k \rightarrow i} \left( x_k \right) 
   \right] \label{eq::marginal1}
\end{align}
and
\begin{align}
  b_{ij} \left( x_i, x_j \right)
  \propto 
  \prod_{l \in L} \prod_{s \in L} \phi_{ik} \left( x_i, x_k \ \middle| \ A; \Gamma \right) m_{i \rightarrow j} \left( x_i \right) m_{j \rightarrow i} \left( x_j \right),
\end{align}
respectively, by using the new messages.

From here, we assume that our observed network is a large sparse graph.
This assumption corresponds to considering $\left| V \right| \gg 1$ and $\Gamma_{ls} = \left. \Gamma_{ls}^\prime \middle/ \left| V \right| \right.$ in our model, where $\Gamma_{ls}^\prime = O \left( 1 \right)$ is a new parameter associated with probability $\Gamma_{ls}$.
We define $\Gamma^\prime = \left\{ \Gamma_{ls}^\prime \ \middle| \ l,s \in L, \Gamma_{ls}^\prime = \Gamma_{sl}^\prime \right\}$ as the set of parameters $\Gamma_{ls}^\prime$.
In this assumption, the messages $\left\{ m_{j \rightarrow i} \left( x_j \right), m_{i \rightarrow j} \left( x_i \right) \ \middle| \ ij \in \overline{E} = I \backslash E \right\}$ can be approximately written as
\begin{align}
  m_{j \rightarrow i} \left( x_j \right)
  & = 
  \frac{1}{z_{j \rightarrow i}} \phi_j \left( x_j \ \middle| \ d_j; \bm{\gamma}, \Theta \right)
  \prod_{k \in V \backslash \left\{ i,j \right\}} 
  \left[ 
    \sum_{ x_k } \phi_{jk} \left( x_j, x_k \ \middle| \ A; \Gamma^\prime \right) m_{k \rightarrow j} \left( x_k \right)
    \right] \nonumber \\
  & = 
  \frac{1}{z_{j \rightarrow i}} \phi_j \left( x_j \ \middle| \ d_j; \bm{\gamma}, \Theta \right)
  \frac{
    \prod_{k \in V \backslash \left\{ j \right\}} \left[ \sum_{ x_k } \phi_{jk} \left( x_j, x_k \ \middle| \ A; \Gamma^\prime \right) m_{k \rightarrow j} \left( x_k \right) \right]
  }{
    1 - \frac{1}{ \left| V \right| }\sum_{ x_i } \prod_{l \in L} \prod_{s \in L} \Gamma_{ls}^{\prime \delta \left( x_i, l \right) \delta \left( x_j, s \right)} m_{i \rightarrow j} \left( x_i \right)
  } \nonumber \\
  & = 
  b_j \left( x_j \right) + O \left( \frac{1}{\left| V \right|} \right), \label{eq::ucMessage}
\end{align}
where we used the relation
\begin{align}
  \frac{1}{1 - y} \simeq 1 - y
\end{align}
for small $y$ and we replaced $\Gamma$ with $\Gamma^\prime$ to clarify the parameter dependence of function $\phi_{ij} \left( x_i, x_j \ \middle| \ A \right)$.
Therefore, the messages from the unconnected vertices $j$ can be regarded as the marginal probability distribution at vertices $j$ for large $\left| V \right|$.
By using Eq. (\ref{eq::ucMessage}), the message passing rule in Eq. (\ref{eq::newMessageTmp}) at $ij \in E$ can be approximated as
\begin{align}
  m_{j \rightarrow i} \left( x_j \right)
  & = 
  \frac{1}{z_{j \rightarrow i}} \phi_j \left( x_j \ \middle| \ d_j; \bm{\gamma}, \Theta \right)
  \prod_{k \in V \backslash \left\{ i,j \right\}} 
  \left[ 
    \sum_{ x_k } \phi_{jk} \left( x_j, x_k \ \middle| \ A; \Gamma^\prime \right) m_{k \rightarrow j} \left( x_k \right)
    \right] \nonumber \\
  & \simeq 
  \frac{1}{z_{j \rightarrow i}^\prime} \phi_j \left( x_j \ \middle| \ d_j; \bm{\gamma}, \Theta \right)
  \prod_{k \in \partial j \backslash \left\{ i \right\}} 
  \left[ 
    \sum_{ x_k } \prod_{l \in L} \prod_{s \in L} \Gamma_{ls}^{\prime \delta \left( x_j, l \right) \delta \left( x_k, s \right)} m_{k \rightarrow j} \left( x_k \right)
    \right] \nonumber \\
  & \quad \times \prod_{k \in \overline{\partial j}}
  \left[
    1 - \frac{1}{\left| V \right|}\sum_{ x_k } \prod_{l \in L} \prod_{s \in L} \Gamma_{ls}^{\prime \delta \left( x_j, l \right) \delta \left( x_k, s \right)} b_k \left( x_k \right)
    \right] \nonumber \\
  & \simeq 
  \frac{1}{z_{j \rightarrow i}^\prime} \phi_j \left( x_j \ \middle| \ d_j; \bm{\gamma}, \Theta \right)
  \exp \left[ - \frac{1}{ \left| V \right| } \sum_{k \in \overline{\partial j}} \sum_{ x_k } \prod_{l \in L} \prod_{s \in L} \Gamma_{ls}^{\prime \delta \left( x_j, l \right) \delta \left( x_k, s \right)} b_k \left( x_k \right) \right] \nonumber \\
  & \quad \times \prod_{k \in \partial j \backslash \left\{ i \right\}} 
  \left[ 
    \sum_{ x_k } \prod_{l \in L} \prod_{s \in L} \Gamma_{ls}^{\prime \delta \left( x_j, l \right) \delta \left( x_k, s \right)} m_{k \rightarrow j} \left( x_k \right)
    \right], \label{eq::BPupdate}
\end{align} 
where $\partial i = \left\{ k \in V \ \middle| \ ik \in E \right\}$, $\overline{\partial j} = \left\{ k \in V \ \middle| \ jk \in \overline{E} \right\}$ and we used the relation
\begin{align}
  \log \left( 1 - y \right) \simeq -y \label{eq::tmptmpapp}
\end{align}
for small $y$.
$z_{j \rightarrow i}^\prime = z_{j \rightarrow i} \left| V \right|^{\left| \partial j \right| - 1}$ is a normalization constant.
Similarly, the approximate marginal distributions $b_i \left( x_i \right)$ and $b_{ij} \left( x_i, x_j \right)$ are approximated as
\begin{align}
  b_i \left( x_i \right)
  & \simeq 
  \frac{1}{z_i^\prime} \phi_i \left( x_i \ \middle| \ d_i; \bm{\gamma}, \Theta \right)
  \exp \left[ - \frac{1}{\left| V \right|}\sum_{k \in \overline{\partial i}} \sum_{ x_k } \prod_{l \in L} \prod_{s \in L} \Gamma_{ls}^{\prime \delta \left( x_i, l \right) \delta \left( x_k, s \right)} b_k \left( x_k \right) \right] \nonumber \\
  & \quad \times \prod_{k \in \partial i } 
  \left[ 
    \sum_{ x_k } \prod_{l \in L} \prod_{s \in L} \Gamma_{ls}^{\prime \delta \left( x_i, l \right) \delta \left( x_k, s \right)} m_{k \rightarrow i} \left( x_k \right)
    \right] \label{eq::appVerBelief}
\end{align}
and
\begin{align}
  b_{ij} \left( x_i, x_j \right)
  & \simeq
  \begin{cases}
    {\displaystyle \frac{1}{z_{ij}^\prime} \prod_{l \in L} \prod_{s \in L} \Gamma_{ls}^{\prime \delta \left( x_i, l \right) \delta \left( x_k, s \right)} m_{i \rightarrow j} \left( x_i \right) m_{j \rightarrow i} \left( x_j \right)}, & ij \in E \\[8pt]
    b_i \left( x_i \right) b_j \left( x_j \right), & ij \in \overline{E} \label{eq::appEdgeBelief}
  \end{cases},
\end{align}
respectively, where $z_i$ and $z_{ij}$ are normalization constants.
It should be noted that the exponential terms in Eqs. (\ref{eq::BPupdate}) and (\ref{eq::appVerBelief}) can be computed by
\begin{align}
  & - \sum_{k \in \overline{\partial i}} \sum_{ x_k } \prod_{l \in L} \prod_{s \in L} \Gamma_{ls}^{\prime \delta \left( x_i, l \right) \delta \left( x_k, s \right)} b_k \left( x_k \right) \nonumber \\
  & = 
  - \sum_{k \in V} \sum_{ x_k } \prod_{l \in L} \prod_{s \in L} \Gamma_{ls}^{\prime \delta \left( x_i, l \right) \delta \left( x_k, s \right)} b_k \left( x_k \right)
  + \sum_{k \in \partial i} \sum_{ x_k } \prod_{l \in L} \prod_{s \in L} \Gamma_{ls}^{\prime \delta \left( x_i, l \right) \delta \left( x_k, s \right)} b_k \left( x_k \right) \label{eq::tmptmptmp}
\end{align}
if we compute the first term of the right hand side in advance before stating the message updates.
Therefore, the computation costs to update each message $m_{j \rightarrow i} \left( x_j \right)$ and marginal distribution $b_i \left( x_i \right)$ are $O \left( \left| \partial j \right| \right)$ and $O \left( \left| \partial i \right| \right)$, respectively.
Because $\left| \partial i \right| \ll \left| V \right|$ for most of the vertices $i \in V$ in many networks, we can compute most of the messages efficiently by using Eq. (\ref{eq::tmptmptmp}).
%The worst case computation time of our method is $O \left( \left| V \right|^2 \right)$ because total message update cost is roughly $\sum_{i \in V} \left| \partial i \right|^2$ and $\left| \partial i \right| \leq \left| V \right| - 1$ in general.
In our method, community labels are estimated by Eqs. (\ref{eq::MPM}) and (\ref{eq::appVerBelief}) after convergence of the new messages using the update rule in Eqs. (\ref{eq::BPupdate}) and (\ref{eq::appVerBelief}).

%% file: EM.tex
\section{Parameter Estimation using EM Algorithm} \label{sec::EM}
In the preceding section, we proposed a method to infer the community labels from the network structure and vertex attribute data based on belief propagation.
However, we have not yet mentioned how to determine the model parameters $\bm{\gamma}, \Gamma^\prime$, and $\Theta$.
It is obvious that community estimation results depend on these parameters.
Therefore, a method that finds the optimal values of these parameters for a given network structure and vertex attribute data is required.
The EM algorithm\cite{DempsterLairdRubinJRSS1977} is one such method that infers the maximum likelihood estimates
\begin{align}
  \left( \widehat{\bm{\gamma}}, \widehat{\Gamma}^\prime, \widehat{\Theta} \right)
  = 
  \arg \max_{ \bm{\gamma}, \Gamma, \Theta } \sum_{ \bm{x} } P \left( \bm{x}, A, \bm{d}; \bm{\gamma}, \Gamma^\prime, \Theta \right) \label{eq::maxlikelihood}
\end{align}
by an iteration method.

In the framework of the EM algorithm, the parameters $\bm{\gamma}, \Gamma^\prime$, and $\Theta$ are estimated by iterative maximization of the $Q$ function.
At iteration $t$, the $Q$ function is written as
\begin{align}
  Q \left( \bm{\gamma}, \Gamma^\prime, \Theta; \bm{\gamma}^{\left( t \right)}, \Gamma^{\prime \left( t \right)}, \Theta^{\left( t \right)} \right)
  & = 
  \sum_{ \bm{x} } P \left( \bm{x} | A, \bm{d}; \bm{\gamma}^{\left( t \right)}, \Gamma^{\prime \left( t \right)}, \Theta^{\left( t \right)} \right) 
  \log P \left( \bm{x}, A, \bm{d}; \bm{\gamma}, \Gamma^\prime, \Theta \right) \nonumber \\
  & =
  \sum_{ i \in V } \sum_{ l \in L } \left< \delta \left( x_i, l \right) \right>^{\left( t \right)}_{\mbox{\scriptsize post}}
  \log \gamma_l \mathcal{N} \left( d_i; \mu_l, \sigma_l \right) \nonumber \\
  & \quad + \sum_{ ij \in E } \sum_{ l \in L } \sum_{ s \in L } \left< \delta \left( x_i, l \right) \delta \left( x_j, s \right) \right>^{\left( t \right)}_{\mbox{\scriptsize post}} \log \Gamma_{ls}^\prime \nonumber \\
  & \quad + \sum_{ ij \in \overline{E} } \sum_{ l \in L } \sum_{ s \in L } \left< \delta \left( x_i, l \right) \delta \left( x_j, s \right) \right>^{\left( t \right)}_{\mbox{\scriptsize post}} \log \left( 1 - \frac{\Gamma_{ls}^\prime}{\left| V \right|} \right) + \mbox{Const.}, \label{eq::Qfunc}
\end{align}
where $\left< f \left( \bm{x} \right) \right>^{\left( t \right)}_{\mbox{\scriptsize post}} = \sum_{ \bm{x} } f \left( \bm{x} \right) P \left( \bm{x} | A, \bm{d}; \bm{\gamma}^{\left( t \right)}, \Gamma^{\prime \left( t \right)}, \Theta^{\left( t \right)} \right)$.
The parameter update rule at iteration $t$ is given by
\begin{align}
  \left( \bm{\gamma}^{\left( t+1 \right)}, \Gamma^{\prime \left( t+1 \right)}, \Theta^{\left( t+1 \right)} \right)
  = 
  \arg \max_{ \bm{\gamma}, \Gamma, \Theta } Q \left( \bm{\gamma}, \Gamma^\prime, \Theta; \bm{\gamma}^{\left( t \right)}, \Gamma^{\prime \left( t \right)}, \Theta^{\left( t \right)} \right).
\end{align}
The maximum likelihood estimates in Eq. (\ref{eq::maxlikelihood}) are given as the convergence point of the above iterative estimation.

By using the belief propagation described in the previous section, we can approximate the expectations in Eq. (\ref{eq::Qfunc}) as
\begin{align}
  \left< \delta \left( x_i, l \right) \right>^{\left( t \right)}_{\mbox{\scriptsize post}}
  = 
  b_i^{\left( t \right)} \left( l \right)
\end{align}
and
\begin{align}
  \left< \delta \left( x_i, l \right) \delta \left( x_j, s \right) \right>^{\left( t \right)}_{\mbox{\scriptsize post}}
  = 
  b_{ij}^{\left( t \right)} \left( l, s \right),
\end{align}
where $b_i^{\left( t \right)} \left( l \right)$ and $b_{ij}^{\left( t \right)} \left( l, s \right)$ are the approximate marginal probability distribution of the posterior probability distribution $P \left( \bm{x} | A, \bm{d}; \bm{\gamma}^{\left( t \right)}, \Gamma^{\prime \left( t \right)}, \Theta^{\left( t \right)} \right)$ computed using Eqs. (\ref{eq::appVerBelief}) and (\ref{eq::appEdgeBelief}), respectively.
Therefore, the $Q$ function in Eq. (\ref{eq::Qfunc}) can be approximated as
\begin{align}
  Q \left( \bm{\gamma}, \Gamma, \Theta; \bm{\gamma}^{\left( t \right)}, \Gamma^{\left( t \right)}, \Theta^{\left( t \right)} \right)
  & \simeq 
  \sum_{ i \in V } \sum_{ l \in L } b_i^{\left( t \right)} \left( l \right)
  \log \gamma_l \mathcal{N} \left( d_i; \mu_l, \sigma_l \right) \nonumber \\
  & \quad + \sum_{ ij \in E } \sum_{ l \in L } \sum_{ s \in L } b_{ij}^{\left( t \right)} \left( l, s \right) \log \Gamma_{ls}^\prime \nonumber \\
  & \quad - \frac{1}{\left| V \right|}\sum_{ ij \in \overline{E} } \sum_{ l \in L } \sum_{ s \in L } b_i^{\left( t \right)} \left( l \right) b_j^{\left( t \right)} \left( s \right) \Gamma_{ls}^\prime + \mbox{Const.} 
\label{eq::appQfunc}
\end{align}
for large $\left| V \right|$ and we used the approximation in Eq. (\ref{eq::tmptmpapp}).
By considering the extreme condition of model parameters $\bm{\gamma}, \Gamma^\prime$, and $\Theta$ in Eq. (\ref{eq::appQfunc}) subject to the constraint in Eq. (\ref{eq::gamma_normalize}), we can find the parameter update rules for each of the parameters are given as
\begin{align}
  \gamma_l^{\left( t+1 \right)}
  & = \frac{1}{\left| V \right|} \sum_{ i \in V } b_i^{\left( t \right)} \left( l \right), \label{eq::EMgamma} \\
  \Gamma_{ls}^{\prime \left( t+1 \right)}
  & = 
  \begin{cases}
    {\displaystyle \frac{2}{\left| V \right| \left( \gamma_l^{ \left( t+1 \right)} \right)^2} \sum_{ ij \in E } b_{ij}^{\left( t \right)} \left( l, s \right)}, & l = s \\
    {\displaystyle \frac{1}{\left| V \right| \gamma_l^{ \left( t+1 \right)} \gamma_s^{ \left( t+1 \right)}} \sum_{ij \in E} \left[ b_{ij}^{\left( t \right)} \left( l, s \right) + b_{ij}^{\left( t \right)} \left( s, l \right) \right]}, & l \not = s
  \end{cases}, \label{eq::EMGamma} \\
  \mu_l^{\left( t+1 \right)}
  & = \frac{ \sum_{ i \in V } d_i b_i^{\left( t \right)} \left( l \right) }{ \sum_{ i \in V } b_i^{\left( t \right)} \left( l \right) }, \label{eq::EMmu}
\end{align}
and
\begin{align}
  \sigma_l^{\left( t+1 \right)}
  = \sqrt{
    \frac{ \sum_{ i \in V } \left( d_i - \mu_l^{\left( t+1 \right)} \right)^2 b_i^{\left( t \right)} \left( l \right) }{ \sum_{ i \in V } b_i^{\left( t \right)} \left( l \right) }
    }, \label{eq::EMsigma}
\end{align}
respectively, where we used the relations
\begin{align}
  \sum_{ ij \in \overline{E} } \left[ b_i^{\left( t \right)} \left( l \right) b_j^{\left( t \right)} \left( s \right) + b_i^{\left( t \right)} \left( s \right) b_j^{\left( t \right)} \left( l \right) \right]
  = &
  \left[ \sum_{ i \in V } b_i^{\left( t \right)} \left( l \right) \right] \left[ \sum_{ j \in V } b_j^{\left( t \right)} \left( s \right) \right] 
  - O \left( \left| V \right| \right),\\
  \sum_{ ij \in \overline{E} } \left[ b_i^{\left( t \right)} \left( l \right) b_j^{\left( t \right)} \left( l \right) \right]
  = &
  \frac{1}{2} \left[ \sum_{ i \in V } b_i^{\left( t \right)} \left( l \right) \right]^2
  - O \left( \left| V \right| \right)
\end{align}
and omitted the ignorable terms for large $\left| V \right|$ to derive Eq. (\ref{eq::EMGamma}).

The proposed algorithm for finding community labels from network structure $A$ and vertex attribute data $\bm{d}$ is summarized as Algorithm \ref{alg::proposed}.
It should be noted that we do not need to compute the messages until convergence at line \ref{alg::line::message} in this algorithm, because it is empirically known that this truncation frequently facilitates the convergence of the EM algorithm faster than one that waits for the convergence of the messages\cite{InoueTanakaFIT2007}.
Further, the fix point of this algorithm also satisfies the message passing equations in Eqs. (\ref{eq::BPupdate}).
Therefore, we compute the convergence point of both the belief propagation and the EM algorithm together in our algorithm.
In addition, we need not compute the approximate marginal probability distributions $b_i \left( x_i \right)$ at line \ref{alg::line::MPM} in our algorithm because these have already been computed at line \ref{alg::line::belief}.

\begin{algorithm}[t]
  \caption{Proposed algorithm} \label{alg::proposed}
  \begin{algorithmic}[1]
    \State Input the adjacency matrix $A$ and the vertex attribute data $\bm{d}$
    \State Initialize all the messages $m_{j \rightarrow i} \left( x_j \right)$, approximate marginal probability distributions $b_i \left( x_i \right)$ and model parameters $\bm{\gamma}, \Gamma^\prime$, and $\Theta$
    \While{no convergence}
    \ForAll{$j \rightarrow i \ \left( i,j \in V, ij \in E \right)$}
    \State update $m_{ji} \left( x_j \right)$ according to Eq. (\ref{eq::BPupdate}) \label{alg::line::message}
    \EndFor
    \ForAll{$i \in V$ and $ij \in E$}
    \State compute $b_i \left( x_i \right)$ according to Eq. (\ref{eq::appVerBelief}) \label{alg::line::belief}
    \State compute $b_{ij} \left( x_i, x_j \right)$ according to Eq. (\ref{eq::appEdgeBelief})
    \EndFor
    \ForAll{$l \in L$ and $\left\{ l,s\right\} \in L^2$}
    \State update $\gamma_l$ according to Eq. (\ref{eq::EMgamma})
    \State update $\Gamma_{ls}^\prime$ according to Eq. (\ref{eq::EMGamma})
    \State update $\mu_l$ according to Eq. (\ref{eq::EMmu})
    \State update $\sigma_l$ according to Eq. (\ref{eq::EMsigma}) \label{alg::line::sigma}
    \EndFor
    \EndWhile
    \State Determine labels $\hat{ \bm{x} }$ according to Eq. (\ref{eq::MPM}) \label{alg::line::MPM}
  \end{algorithmic}
\end{algorithm}

%% file: exp.tex
\section {Numerical Experiment} \label{sec::Exp}
\begin{figure}[t]
  \begin{center}
    \includegraphics[width=7.0cm]{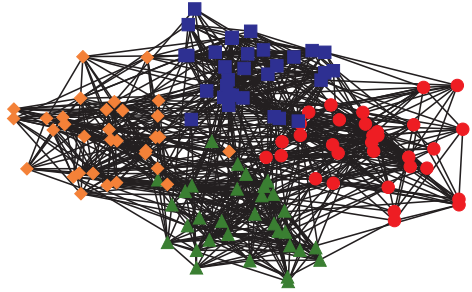}
    \caption{
      Example of a computer-generated network with 128 vertices, 1066 edges, and 4 communities.
      We set $p_{\mbox{\scriptsize in}} = 13.5 / 31$ and $p_{\mbox{\scriptsize out}} = 2.5 / 96$ to create this network.
      The modularity of this network is 0.504.
    }\label{fig::compNet}
  \end{center}
\end{figure}

\begin{figure}[t]
  \begin{center}
    \includegraphics[width=5.5cm]{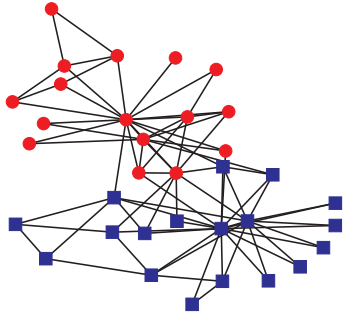}
    \caption{
      Zachary's karate club network with 34 vertices, 78 edges, and 2 communities.
      This network represents the friendship between the members of a karate club.
      Each community in this network expresses the factions of the club.
      The modularity of this network is 0.371.
    }\label{fig::karate}
  \end{center}
\end{figure}

\begin{figure}[t]
  \begin{center}
    \includegraphics[width=5.5cm]{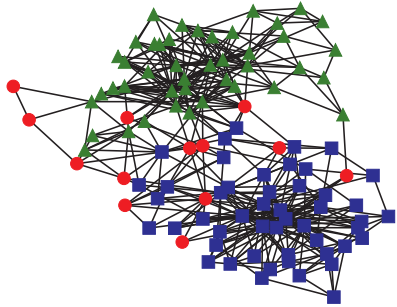}
    \caption{
      Network of the books about US politics created by Krebs with 105 vertices, 441 edges, and 3 communities.
      This network represents the co-purchasing relationship of the books sold by the online bookseller Amazon.com.
      Each community in this network expresses the principles of each book (``liberal,'' ``neutral,'' and ``conservative'').
      The modularity of this network is 0.415.
    }\label{fig::polbooks}
  \end{center}
\end{figure}

\begin{figure}[t]
  \begin{center}
    \includegraphics[width=7.0cm]{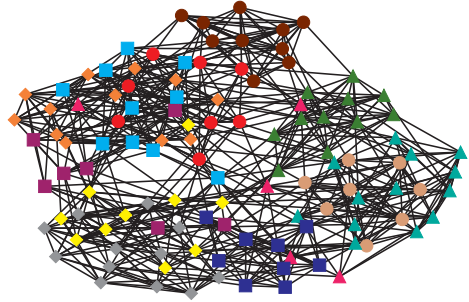}
    \caption{
      Network of American football games between Division IA colleges during the regular season in Fall 2000 with 115 vertices, 613 edges, and 12 communities.
      The vertices and edges represent the colleges and games between teams, respectively.
      The communities in this network represent the conferences to which they belong.
      The modularity of this network is 0.554.
    }\label{fig::football}
  \end{center}
\end{figure}

In this section, we describe the numerical verification of the performance of our method.
We used computer-generated networks composed of $128$ vertices separated into four communities of the same size and real-world networks created manually in these experiments.
The computer-generated networks were samples drawn from the stochastic block model in Eq. (\ref{eq::SBM}).
They correspond to the benchmark networks called the four group test used by Girvan and Newman\cite{GirvanNewmanNAS2002, NewmanGirvanPRE2004}, where the edges are locations between the pairs of vertices belonging to the same community with probability $p_{\mbox{\scriptsize in}}$, while pairs of vertices belonging to different communities are linked with probability $p_{\mbox{\scriptsize out}}$.
The values of $p_{\mbox{\scriptsize in}}$ and $p_{\mbox{\scriptsize out}}$ are chosen to satisfy
\begin{align}
  31 p_{\mbox{\scriptsize in}} + 96 p_{\mbox{\scriptsize out}} = 16
\end{align}
so that the expected degree of each vertex equals $16$.
The real-world networks used in these experiments were a karate club network\cite{ZacharyJAnthrR1977}, books about US politics\cite{KrebsUP}, and an American football games network\cite{GirvanNewmanNAS2002}.
The true community labels in these networks were determined by its creators.
Examples of the computer-generated and real-world networks are shown in Figs. \ref{fig::compNet}-\ref{fig::football}, where the true communities are represented by the vertex colors and shape.
The vertex attribute data were generated according to the conditional probability density function in Eq. (\ref{eq::condi2}), where the true mean values and variance were set as $\mu_1 = 0, \mu_2 = 10, \dots, \mu_{L_{\mbox{\scriptsize max}}} = 10 \left( L_{\mbox{\scriptsize max}} - 1 \right) $ and $\sigma_1 = \sigma_2 = \cdots = \sigma_{L_{\mbox{\scriptsize max}}} = \sigma$, respectively.
The performances of our algorithm were evaluated for various $p_{\mbox{\scriptsize out}}$ and values of $\sigma$ in these experiments.


We evaluated the performances of our method by the average accuracy and average modularity over 500 trials defined as
\begin{align}
  \left[ \mbox{Accuracy} \right]
  = 
  \frac{1}{500} \sum_{t = 1}^{500} \mbox{Accuracy} \left( \hat{ \bm{x} }_t, \bm{x}^* \right)
\end{align}
and
\begin{align}
  \left[ \mbox{Modularity} \right]
  = 
  \frac{1}{500} \sum_{t = 1}^{500} \mbox{Modularity} \left( \hat{ \bm{x} } \right),
\end{align}
respectively.
$\mbox{Accuracy} \left( \hat{ \bm{x} }_t, \bm{x}^* \right)$ and $\mbox{Modularity} \left( \hat{ \bm{x} } \right)$ are defined as
\begin{align}
  \mbox{Accuracy} \left( \hat{ \bm{x} }, \bm{x}^* \right)
  = 
  \max_{ \rho } \frac{1}{\left| V \right|} \sum_{ i \in V } \delta \left( \rho \left( \hat{ x }_i \right), x_i^* \right)
\end{align}
and 
\begin{align}
  \mbox{Modularity} \left( \hat{ \bm{x} } \right)
  = 
  \frac{1}{2 \left| E \right|} \sum_{ i \in V } \sum_{ j \in V } \left( A_{ij} - \frac{ \left| \partial i \right| \left| \partial j \right| }{ 2 \left| E \right| } \right) \delta \left( x_i, x_j \right),
\end{align}
where 
$\bm{x}^* = \left\{ x_i^* \in L \middle| i \in V \right)$ is a set of true community labels and $\rho$ ranges over the permutation on $L_{\mbox{\scriptsize max}}$ elements.
Empirically, it is said that a network divided by assigning the labels has a community structure if its modularity is greater than $0.3$\cite{NewmanPRE2004,ClausetNewmanMoorePRE2004}.
For each trial, the vertex attribute data (for both cases) and the network structure (for only computer-generated network case) were newly generated.
We compared our method with three different types of competitive method, a method that considers both network structure and vertex attribute data and a method that utilizes either type of information.
The first method was the naive mean field method proposed by Zanghi et al.\cite{ZanghiVolantAmbroisePattRecogLett2010}.
The second type is methods that utilize only network structures.
We chose the Newman method and the message passing method proposed by Decelle et al.\cite{DecelleKrzakalaMooreZdeborovaPRL2011, DecelleKrzakalaMooreZdeborovaPRE2011} as the competitors of this type.
The last type is the data clustering method that considers only vertex attribute data;the k-means++ algorithm\cite{ArthurVassilvitskiiACM-SIAM2007} was used for comparison with our method.

\begin{figure*}[t]
  \begin{tabular}{c c}
    (a) & (b) \\
    \includegraphics[width=8.0cm]{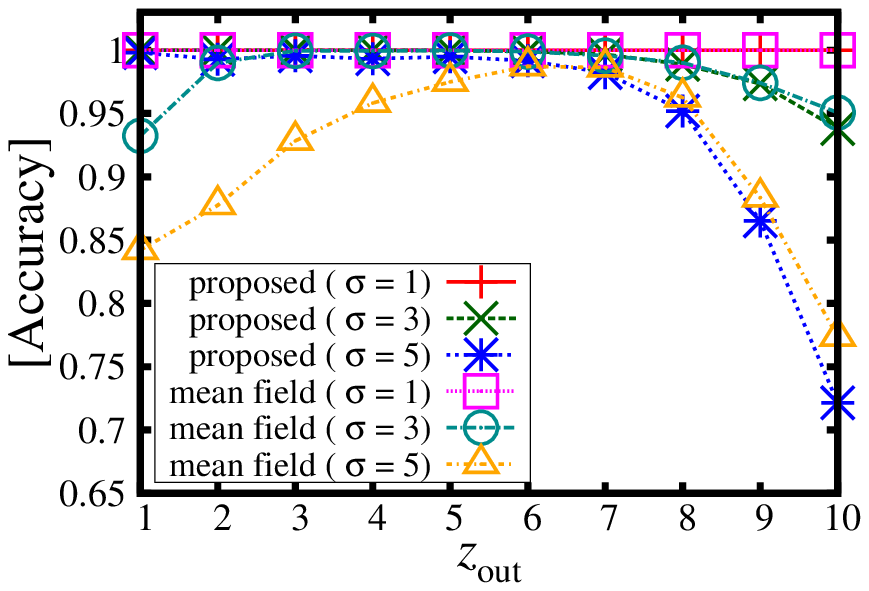}
    &
    \includegraphics[width=8.0cm]{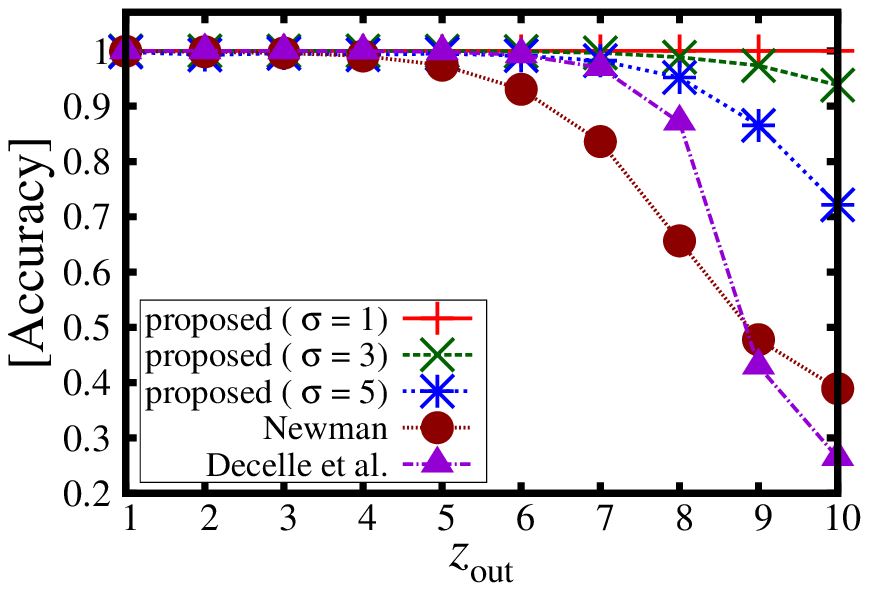}
    \\
    (c) & (d) \\
    \includegraphics[width=8.0cm]{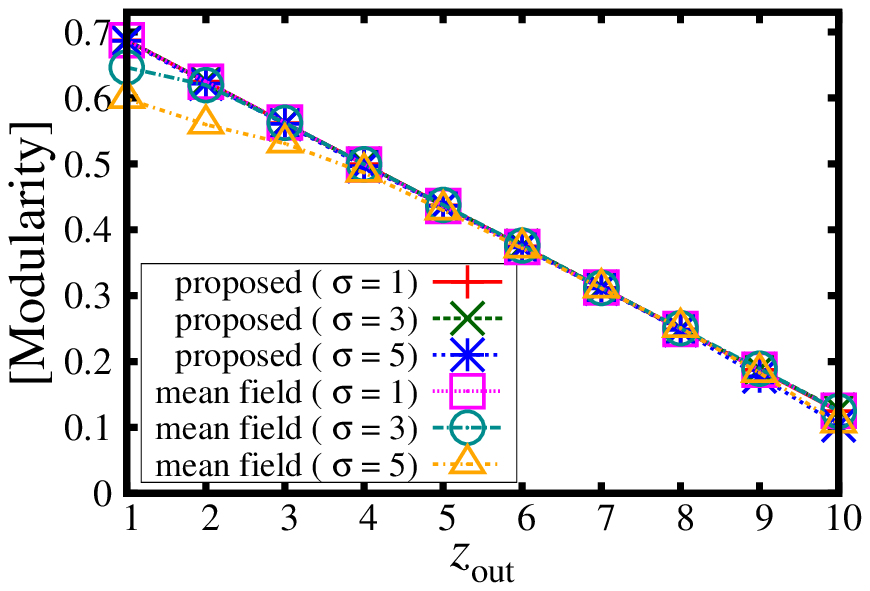}
    &
    \includegraphics[width=8.0cm]{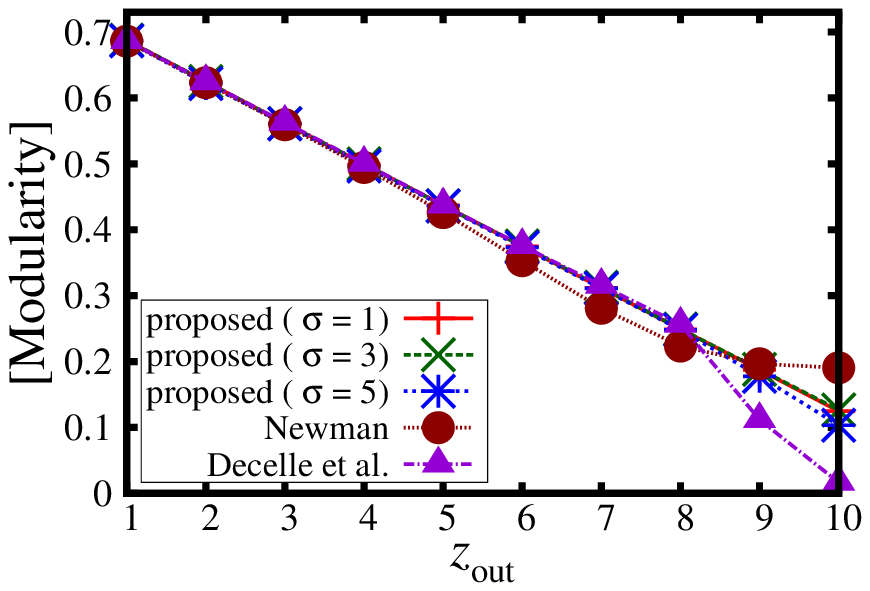}
  \end{tabular}
  \caption{
    Comparisons of $\left[ \mbox{Accuracy} \right]$ and $\left[ \mbox{Modularity} \right]$ versus $z_{\mbox{\scriptsize out}}$ in the case of the computer-generated network.
    Each point is obtained by averaging over $500$ trials.
    (a) $\left[ \mbox{Accuracy} \right]$ versus $z_{\mbox{\scriptsize out}}$ of our method and the naive mean field method for $\sigma = 1$, $\sigma = 3$, $\sigma = 5$.
    (b) $\left[ \mbox{Accuracy} \right]$ versus $z_{\mbox{\scriptsize out}}$ of our method, the Newman method, and the message passing method proposed by Decelle et al. for $\sigma = 1$, $\sigma = 3$, $\sigma = 5$
    (c) $\left[ \mbox{Modularity} \right]$ versus $z_{\mbox{\scriptsize out}}$ of our method and the naive mean field method for $\sigma = 1$, $\sigma = 3$, $\sigma = 5$.
    (d) $\left[ \mbox{Modularity} \right]$ versus $z_{\mbox{\scriptsize out}}$ of our method, the Newman method, and the message passing method proposed by Decelle et al. for $\sigma = 1$, $\sigma = 3$, $\sigma = 5$
  }\label{fig::compNet_numOut}
\end{figure*}
\begin{table*}
  \caption{
    Average modularities computed from $500$ samples drawn from the stochastic block model
    } \label{table::comNet}
  \begin{tabular}{ | c | c | c | c | c | c | c | c | c | c | c | } \hline
    $z_{\mbox{\scriptsize out}}$ & 1 & 2 & 3 & 4 & 5 & 6 & 7 & 8 & 9 & 10 \\ \hline
    $\left[ \mbox{Modularity} \right]$ & 0.687 & 0.624 & 0.562 & 0.499 & 0.437 & 0.375 & 0.311 & 0.248 & 0.188 & 0.124 \\ \hline 
  \end{tabular}
\end{table*}

In these experiments, we chose the initial values of $m_{ji} \left( x_j \right)$ and approximate marginal probability distributions $b_i \left( x_i \right)$ for solving the message update Eqs. (\ref{eq::BPupdate}) and (\ref{eq::appVerBelief}) and $\bm{\gamma}^{\left( 0 \right)}, \Gamma^{\left( 0 \right)}$, and $\Theta^{\left( 0 \right)}$ for the parameter update rules in Eqs. (\ref{eq::EMgamma})-(\ref{eq::EMsigma}) as follows:
\begin{align}
  m_{j \rightarrow i} \left( x_j \right) 
  = & 
  \frac{u_{ji} \left( x_j \right)}{\sum_{ l \in L } u_{ji} \left( l \right)}, \\
  b_i \left( x_i \right)
  = & 
  \frac{u_{i} \left( x_i \right)}{\sum_{ l \in L } u_{i} \left( l \right)}, \label{eq::init1} \\
  \gamma_l^{\left( 0 \right)}
  = & 
  \frac{1}{L_{\mbox{\scriptsize max}}}, \\
  \Gamma_{ls}^{\prime \left( 0 \right)}
  = & 
  \frac{2 \left| E \right|}{\left| V \right| - 1 }, \\
  \mu_l^{\left( 0 \right)}
  = & c_l,
\end{align}
and
\begin{align}
  \sigma_l^{\left( 0 \right)}
  = 1, \label{eq::init2}
\end{align}
respectively, where $u_{ji} \left( x_j \right)$ and $u_i \left( x_j \right)$ are random numbers drawn from the uniform distributions, the support of which is $[0,1)$, and $c_l$ is a central point obtained by the k-means++ algorithm.

\begin{figure*}[t]
  \begin{tabular}{c c}
    (a) & (b) \\
    \includegraphics[width=8.0cm]{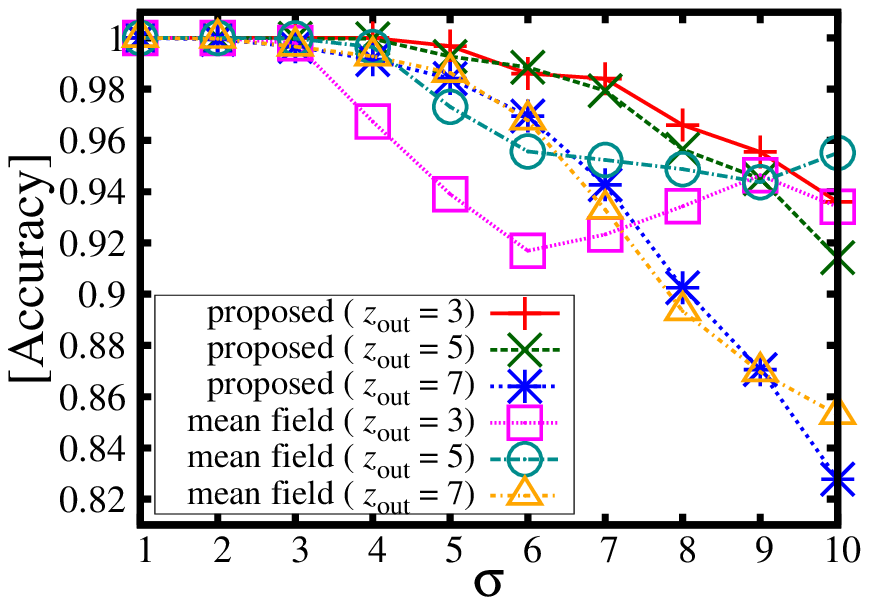}
    &
    \includegraphics[width=8.0cm]{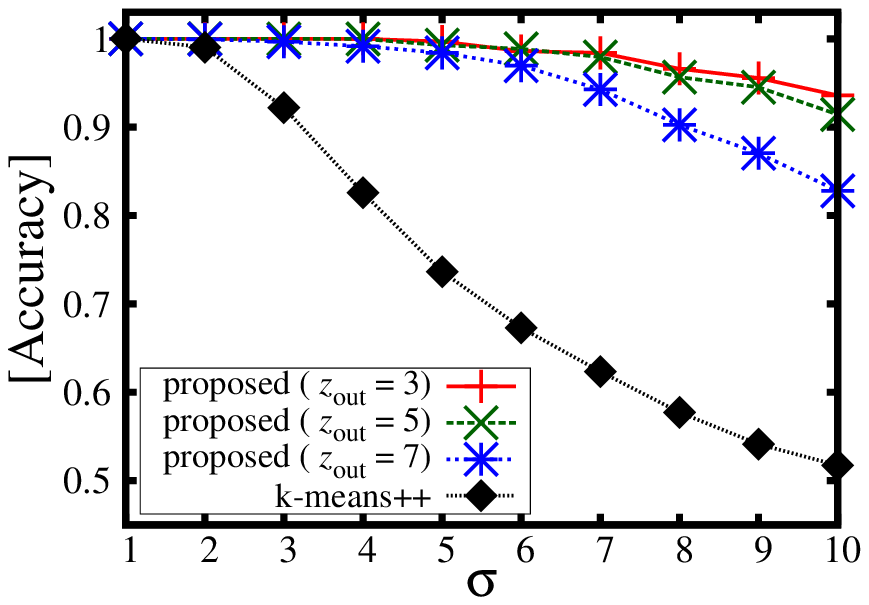}
    \\
    (c) & (d) \\
    \includegraphics[width=8.0cm]{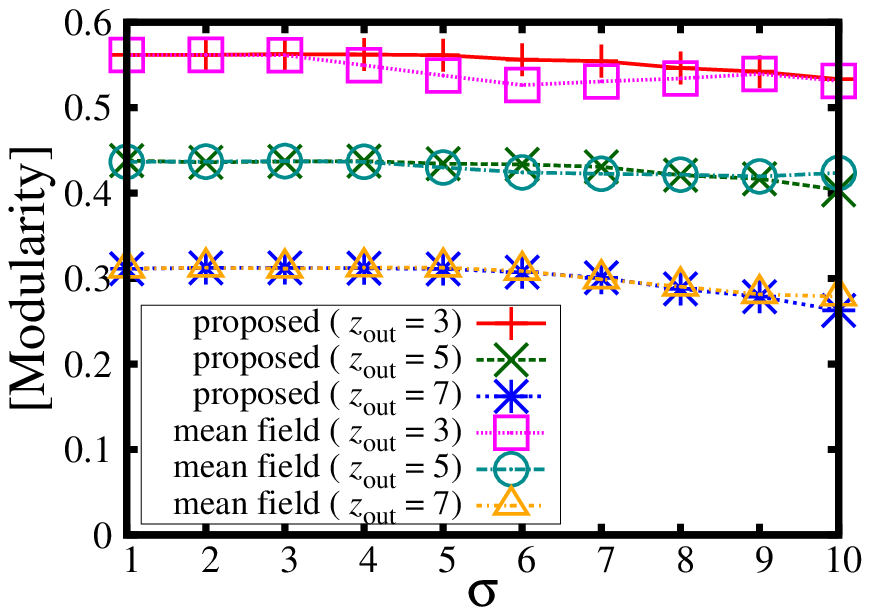}
    &
    \includegraphics[width=8.0cm]{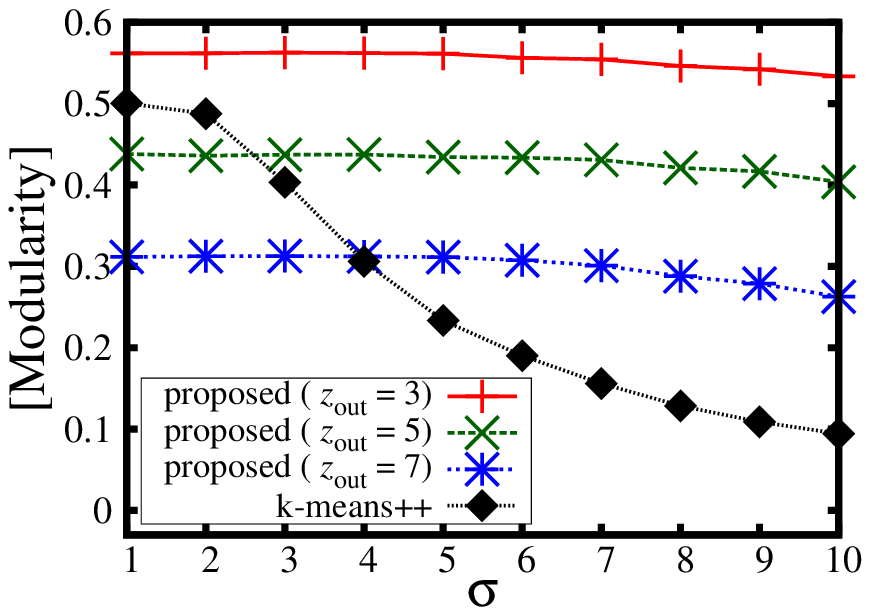}
  \end{tabular}
  \caption{
    Comparisons of $\left[ \mbox{Accuracy} \right]$ and $\left[ \mbox{Modularity} \right]$ versus $\sigma$ in the case of the computer-generated network.
    Each point is obtained by averaging over $500$ trials.
    (a) $\left[ \mbox{Accuracy} \right]$ versus $\sigma$ of our method and the naive mean field method for $z_{\mbox{\scriptsize out}} = 3$, $z_{\mbox{\scriptsize out}} = 5$, $z_{\mbox{\scriptsize out}} = 7$.
    (b) $\left[ \mbox{Accuracy} \right]$ versus $\sigma$ of our method and k-means++ for $z_{\mbox{\scriptsize out}} = 3$, $z_{\mbox{\scriptsize out}} = 5$, $z_{\mbox{\scriptsize out}} = 7$.
    (c) $\left[ \mbox{Modularity} \right]$ versus $\sigma$ of our method and the naive mean field method for $z_{\mbox{\scriptsize out}} = 3$, $z_{\mbox{\scriptsize out}} = 5$, $z_{\mbox{\scriptsize out}} = 7$.
    (d) $\left[ \mbox{Modularity} \right]$ versus $\sigma$ of our method and k-means++ for $z_{\mbox{\scriptsize out}} = 3$, $z_{\mbox{\scriptsize out}} = 5$, $z_{\mbox{\scriptsize out}} = 7$.
  }\label{fig::compNet_noise}
\end{figure*}

Figures \ref{fig::compNet_numOut} and \ref{fig::compNet_noise} show the plot of average accuracy and average modularity versus $z_{\mbox{\scriptsize out}}$ and $\sigma$ for the computer-generated networks, respectively, where $z_{\mbox{\scriptsize out}}$ is an average number of edges connecting different communities per vertex, that is, $z_{\mbox{\scriptsize out}} = 96 p_{\mbox{\scriptsize out}}$.
The setting of the initial values for the naive mean field method was the same as for our method in Eqs. (\ref{eq::init1})-(\ref{eq::init2}) (messages were unused in the naive mean field method and parameter $\sigma_l$ was restricted as $\sigma_1 = \sigma_2 =\cdots = \sigma_{L_{ \mbox{\scriptsize max}} }$ in the original studies of Zanghi et al.).
Figures \ref{fig::compNet_numOut} (a) and (b) show the plot of $\left[ \mbox{Accuracy} \right]$ versus $z_{\mbox{\scriptsize out}}$ and Figs. \ref{fig::compNet_numOut} (c) and (d) show the plot of $\left[ \mbox{Modularity} \right]$ versus $z_{\mbox{\scriptsize out}}$ for $\sigma = 1$, $\sigma = 3$, $\sigma = 5$.
The average modularities of the computer-generated networks for the true community labels are given in Table \ref{table::comNet} and it can be seen that the computer-generated networks have community structures for $z_{\mbox{\scriptsize out}} \leq 7$.
Similarly, Figs. \ref{fig::compNet_noise} (a) and (b) show the plot of $\left[ \mbox{Accuracy} \right]$ versus $\sigma$ and Figs. \ref{fig::compNet_noise} (c) and (d) show the plot of $\left[ \mbox{Modularity} \right]$ versus $\sigma$ for $z_{\mbox{\scriptsize out}} = 3$, $z_{\mbox{\scriptsize out}} = 5$, $z_{\mbox{\scriptsize out}} = 7$.
In Fig. \ref{fig::compNet_numOut}, it can be seen that, while being slightly inferior in average accuracy to the naive mean field method and in average modularity to the Newman method for high $z_{\mbox{\scriptsize out}}$, our method performs better than all the competitive methods in the region where $z_{\mbox{\scriptsize out}}$ is relatively small and the computer-generated network has a community structure.
In Fig. \ref{fig::compNet_noise}, it can be seen that our method also performs better than the other methods in average accuracy in the region where the value of $\sigma$ is small.
Because the interval between the true mean values of the vertex attribute data is $10$, the fraction of the similar attribute values in different communities grows when the value of $\sigma$ is greater than approximately $2.5$.
Therefore, the detection problem becomes difficult when the value of $\sigma$ is greater than $2.5$.
However, the methods that consider both the network structure and vertex attribute data produce high accuracy results (over $80\%$) for a large value of $\sigma$ and are robust to $\sigma$ in the modularity measure.
In the case of computer-generated networks, the performance of our method is superior to that of all the competitive methods when the network has a community structure (corresponding to a modularity of true community labels over 0.3) and the vertex attribute data are well divided to allow detection of the communities (corresponding to a small value of $\sigma$).

\begin{figure*}[t]
  \begin{tabular}{c c}
    (a) & (b) \\
    \includegraphics[width=8.0cm]{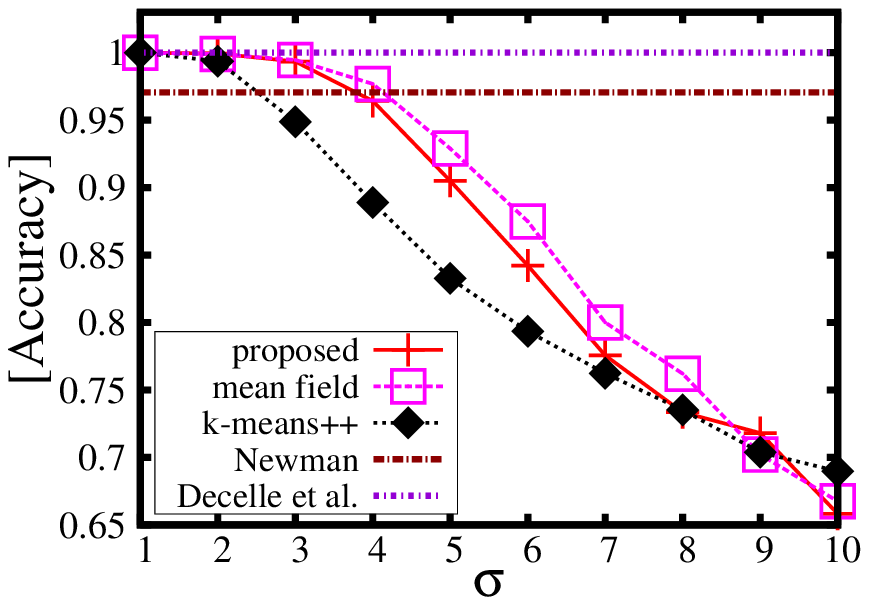}
    &
    \includegraphics[width=8.0cm]{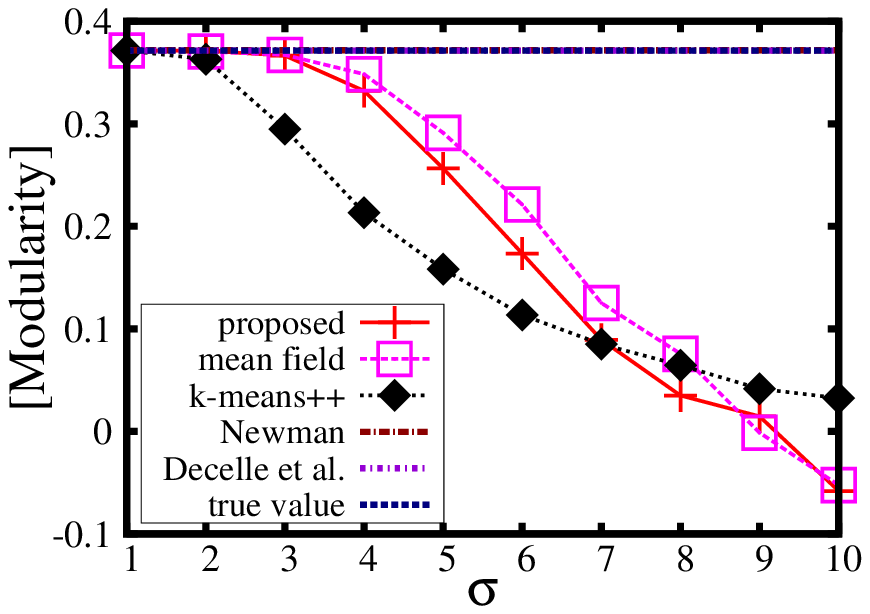}
  \end{tabular}
  \caption{
    Comparisons of $\left[ \mbox{Accuracy} \right]$ and $\left[ \mbox{Modularity} \right]$ versus $\sigma$ in the case of the karate club network.
    (a) $\left[ \mbox{Accuracy} \right]$ versus $\sigma$ of our method and  all the competitive methods.
    The results of the method that considers only network structure are represented as horizontal lines.
    (b)$\left[ \mbox{Modularity} \right]$ versus $\sigma$ of our method and all the competitive methods.
    The results of the method that considers only network structure are represented as horizontal lines together with the true modularity value.
  }\label{fig::karate_noise}
\end{figure*}

\begin{figure*}[t]
  \begin{tabular}{c c}
    (a) & (b) \\
    \includegraphics[width=8.0cm]{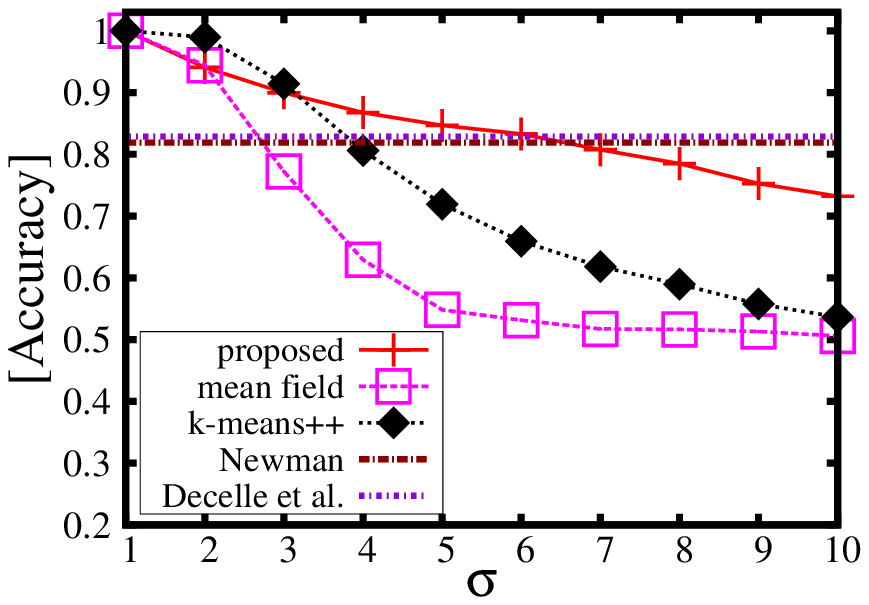}
    &
    \includegraphics[width=8.0cm]{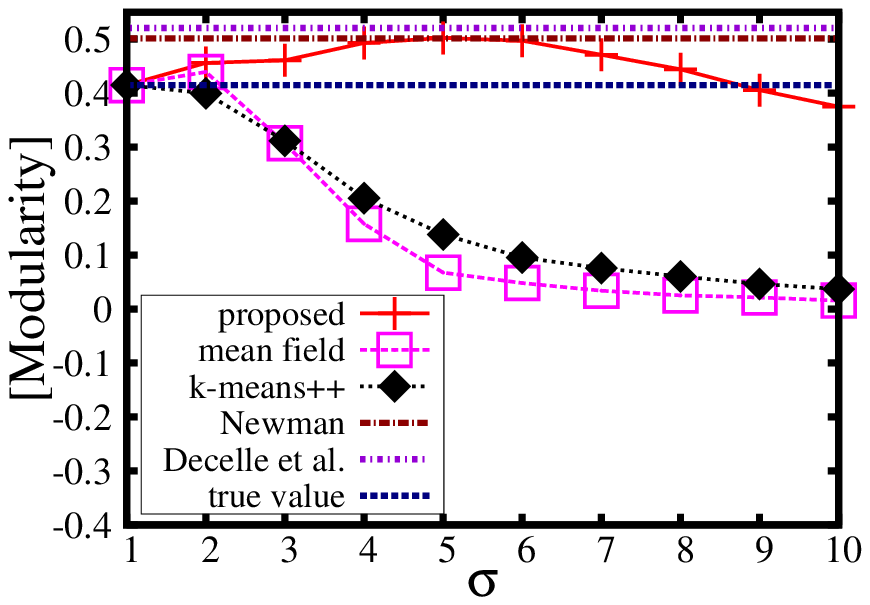}
  \end{tabular}
  \caption{
    Comparisons of $\left[ \mbox{Accuracy} \right]$ and $\left[ \mbox{Modularity} \right]$ versus $\sigma$ in the case of the US politics books network.
    (a) $\left[ \mbox{Accuracy} \right]$ versus $\sigma$ of our method and all the competitive methods.
    The results of the method that considers only network structure are represented as horizontal lines.
    (b)$\left[ \mbox{Modularity} \right]$ versus $\sigma$ of our method and all the competitive methods.
    The results of the method that considers only network structure are represented as horizontal lines together with the true modularity value.
  }\label{fig::polbooks_noise}
\end{figure*}

\begin{figure*}[t]
  \begin{tabular}{c c}
    (a) & (b) \\
    \includegraphics[width=8.0cm]{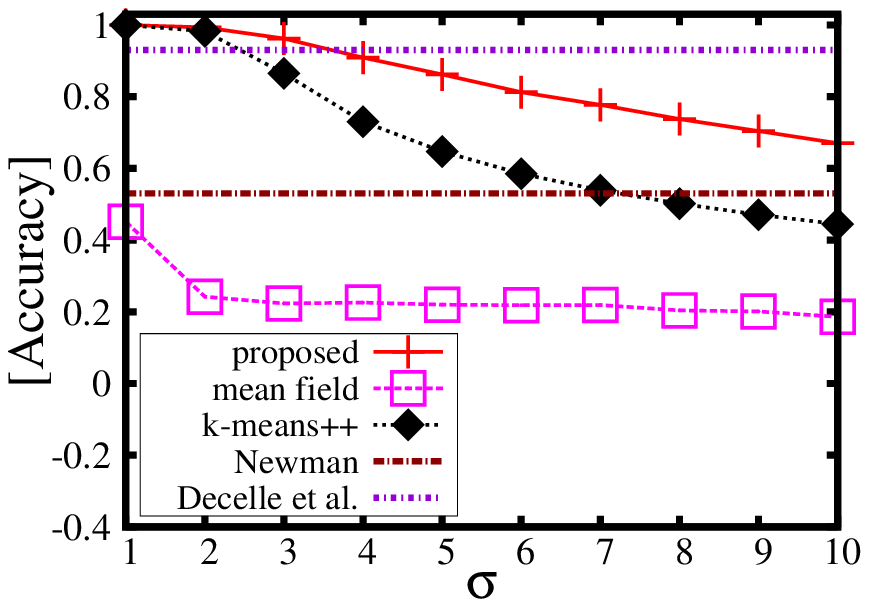}
    &
    \includegraphics[width=8.0cm]{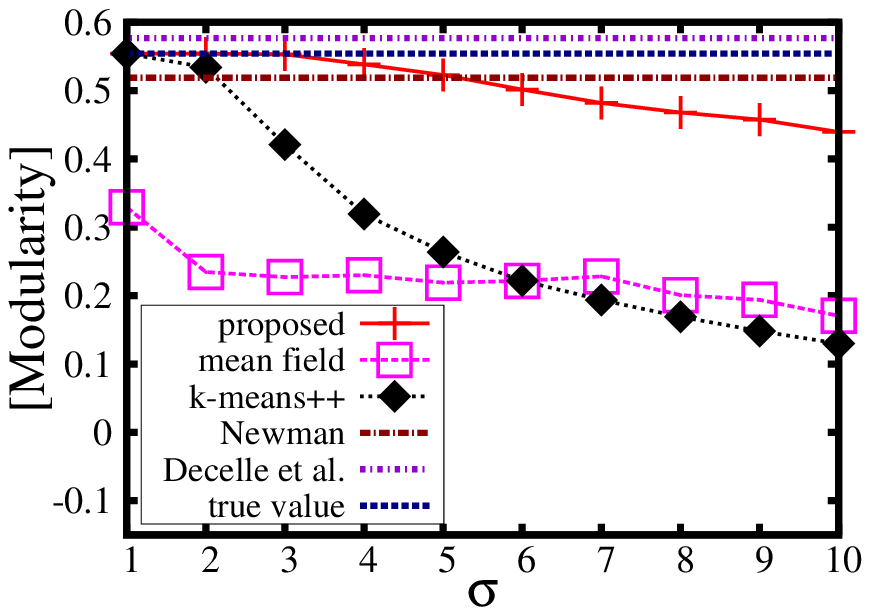}
  \end{tabular}
  \caption{
    Comparisons of $\left[ \mbox{Accuracy} \right]$ and $\left[ \mbox{Modularity} \right]$ versus $\sigma$ in the case of the American football games network.
    (a) $\left[ \mbox{Accuracy} \right]$ versus $\sigma$ of our method and all the competitive methods.
    The results of the method that considers only network structure are represented as horizontal lines.
    (b)$\left[ \mbox{Modularity} \right]$ versus $\sigma$ of our method with all the competitive methods.
    The results of the method that considers only network structure are represented as horizontal lines together with the true modularity value.
  }\label{fig::football_noise}
\end{figure*}

Figures \ref{fig::karate_noise}-\ref{fig::football_noise} show the plot of the average accuracy and average modularity versus $\sigma$ in the case of the real-world networks.
Figure \ref{fig::karate_noise} shows the results for the karate club network.
In this figure, the methods that consider both the network structure and vertex attribute data yield better results than the other competitive methods in the region where the value of $\sigma$ is small.
However, the performance of our method is slightly inferior to that of the naive mean field method proposed by Zanghi et al.
In our opinion, this result has its root in the approximation of our message passing rule in Eqs. (\ref{eq::ucMessage}) and (\ref{eq::BPupdate}) assuming large $\left| V \right|$, because the size of this network is small.

Figure \ref{fig::polbooks_noise} shows the results for the network of books about US politics.
In this figure, it can be seen that our method yields higher accuracy than all the competitive methods without the case where the value of $\sigma$ is small.
We consider that this result stems from the fact that the neutral community represented by the red circle in Fig. \ref{fig::polbooks} has sparse intra-connection.
Therefore, many vertices in this community were considered to be assigned wrong labels for this reason.
In our opinion, this is the reason why k-means++ yields the best accuracy in the region where the value of $\sigma$ is small.
However, in most regions of $\sigma$ our method yields the modularity closest to the true value in this network.
This result means that our method infers the community structure, the connectivity of which is close to the true community structure.

Figure \ref{fig::football_noise} shows the plot for average accuracy and average modularity versus $\sigma$ in the case of the American football games network.
In this figure, it can be seen that our method yields better results than the other competitive methods in the region where the value of $\sigma$ is small.

%% file: conc.tex
\section{Concluding Remarks} \label{sec::conclude}
In this paper, we proposed a new community detection method that considers both the network structure and vertex attribute data.
Our method can be regarded as an extension of the previous method proposed by Zanghi et al.\cite{ZanghiVolantAmbroisePattRecogLett2010} from the perspective of the cluster variation method or as a combination of the message passing method proposed by Decelle et al.\cite{DecelleKrzakalaMooreZdeborovaPRL2011, DecelleKrzakalaMooreZdeborovaPRE2011} and a traditional data clustering method using the mixture of Gaussian distribution.
In our method, the detection of the community labels is reduced to solving a simultaneous equation of the message passing rule of belief propagation.
The model parameters in the posterior probability distribution are determined from the network structure and vertex attribute data by using the EM algorithm.
We evaluated the performance of our method by applying it to computer-generated and real-world networks and by comparing its results with those of several types of competitive detection method in numerical experiments. We verified that our community detection method can infer the community labels with high accuracy if the network has a community structure and the vertex attribute data are sufficiently divided.

In our method, the number of communities $L_{\mbox{\scriptsize max}}$ must be determined in advance and we used the true number of communities for each network in the numerical experiments.
It is ideal to infer $L_{\mbox{\scriptsize max}}$ from the network data.
One possible method is to use the modularity for determining $L_{\mbox{\scriptsize max}}$, as in Newman's method, that is, to conduct our detection method for several $L_{\mbox{\scriptsize max}}$ and adopt the best result that gives the maximum modularity value.
However, this method considers only network structure and ignores the contribution of the vertex attribute data for determining $L_{\mbox{\scriptsize max}}$.
Therefore, we need to seek a further suitable method to estimate the optimal $L_{\mbox{\scriptsize max}}$.

We assumed that the network structure was drawn from the stochastic block model where the structures of each community corresponded to the Erd\H{o}s-R\'enyi random graph\cite{ErdosRenyiPubMath1959} in our scheme.
Other types of stochastic block model exist in the field of complex networks\cite{KarrerNewmanPRE2011} and a message passing algorithm of the degree-corrected block models has already been proposed\cite{YanShaliziJensenKrzakaraMooreZdeborovaZhangZhuJStatMech2014}. 
By virtue of the flexibility of the Bayesian framework adopted in this work, we can extend our method to other types of stochastic block model and create a more suitable model for more realistic networks considering both a more accurate network structure and vertex attribute data.

The other direction in which our model can be extended is to consider further Bayesian treatment.
The combination of our message passing approach and variational Bayesian methods\cite{RyotaOkadaMiyoshiJPSJ2011} is a very interesting extension and will produce a more efficient algorithm to detect community structures.
We aim to develop our method in these directions.